\documentclass{rmf-d}
\usepackage{nopageno,rmfbib,multicol,times,epsf,amsmath,amssymb,cite}
\usepackage[latin1]{inputenc}
\usepackage[]{caption2}
\usepackage{graphicx}
\usepackage{wrapfig}

\def\rmfcornisa{
}

\def\rmfcintilla{
}
\clearpage \rmfcaptionstyle 
\begin{document}
\title{Visualization of internal forces inside the proton in a
  classical relativistic model
\vspace{-6pt}}
\author{Mira Varma}
\address{Physics Department, Yale University, New Haven, CT 06520, USA}
\author{Peter Schweitzer}
\address{Department of Physics, University of Connecticut, Storrs, CT 06269, USA}
\maketitle
\recibido{day month year}{day month year
\vspace{-12pt}}
\begin{abstract}
\vspace{1em} Abstract: 
A classical model of a stable particle of finite size is studied. The model parameters can be chosen such that the described particle has the mass and radius of a proton. Using the energy-momentum tensor (EMT), we show how the presence of long-range forces alters some notions taken for granted in short-range systems. We focus our attention on the $D$-term form factor. The important conclusion is that a more careful definition of the $D$-term may be required when long-range forces are present.   \vspace{1em}
\end{abstract}
\keys{  energy-momentum tensor, $D$-term, classical model  \vspace{-4pt}}
\pacs{   \bf{\textit{03.50.-z, 11.27.+d, 14.20.Dh, 21.30.Fe }}    \vspace{-4pt}}
\begin{multicols}{2}

\section{Introduction}

In this proceeding, we review the results for EMT densities  from Ref.~\cite{Varma:2020crx} based on Bia\l ynicki-Birula's classical model of the proton (BB-model) \cite{Bialynicki-Birula:1993shm}. The EMT can be studied through generalized parton distribution functions in hard exclusive reactions, and is of interest because it contains information about the basic properties of a particle: the mass, spin, and $D$-term \cite{Kobzarev:1962wt,Pagels:1966zza,Ji:1998pc,Radyushkin:2000uy,Goeke:2001tz,Diehl:2003ny,Belitsky:2005qn,Boffi:2007yc}. Although less well-known, the $D$-term is of equal importance as the other basic properties \cite{Polyakov:1999gs}. From the $D$-term form factor $D(t)$ and other EMT form factors, one can learn about the 2D and 3D distributions of energy, angular momentum and internal forces \cite{Polyakov:2002yz,Polyakov:2018zvc,Lorce:2018egm,Freese:2021czn,Panteleeva:2021iip}. 
Recently, first experimental insights on the $D$-term 
became available \cite{Kumano:2017lhr,Burkert:2018bqq,Kumericki:2019ddg,Burkert:2021ith}. 

In Ref.~\cite{Varma:2020crx} the BB-model \cite{Bialynicki-Birula:1993shm} was used to understand the impact of long-range forces not considered in other theoretical studies in systems with exclusively short-range forces \cite{Goeke:2007fp,Goeke:2007fq,Cebulla:2007ei,Jung:2013bya,Kim:2012ts,Jung:2014jja,Mai:2012yc,Mai:2012cx,Cantara:2015sna,Gulamov:2015fya,Nugaev:2019vru,Hudson:2017xug,Hudson:2017oul,Shanahan:2018nnv,Shanahan:2018pib,Anikin:2019kwi,Neubelt:2019sou,Azizi:2019ytx,Panteleeva:2020ejw,Kim:2020nug,Chakrabarti:2020kdc,Owa:2021hnj,More:2021stk}.
The BB-model \cite{Varma:2020crx} yields qualitatively similar results to experimental insights \cite{Burkert:2018bqq,Burkert:2021ith}. 
Even though it is classical, the BB-model is well suited for our purpose since it lets us investigate the impact of long-range forces without worrying about technical difficulties that arise in quantum field theory. We will later show that our conclusions about the impact of the long-range forces are model independent.

\section{EMT Tensor in the Classical Model}

The BB-model \cite{Bialynicki-Birula:1993shm} consists of `` dust particles '' in a spherically symmetric region of radius $R$ bound by three fields: a massive scalar field, $\phi$, a massive vector field, $V^\mu$, and an electromagnetic field, $A^\mu$. The particles couple to these fields via the coupling constants $g_S$, $g_V$, and the electric charge $e$. The classical field equations are relativistic and can be found in Refs. \cite{Varma:2020crx,Bialynicki-Birula:1993shm}. The parameters $g_S$, $m_S$, $g_V$, $m_V$ correspond, respectively, to the coupling constants and masses of sigma and omega mesons as used in nuclear models \cite{Bialynicki-Birula:1993shm}.

\begin{figure*}
\begin{center}
\includegraphics[width=.32\linewidth]{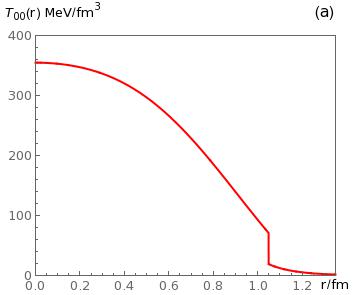} \
\includegraphics[width=.32\linewidth]{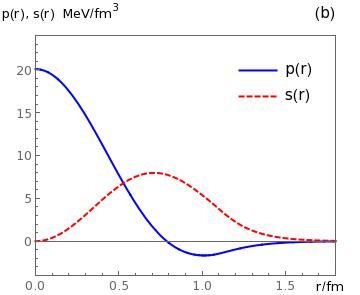} \
\includegraphics[width=.32\linewidth]{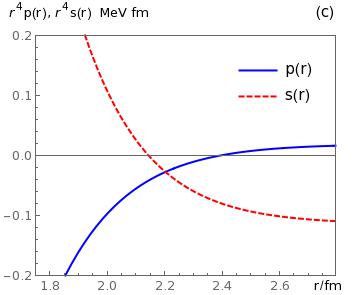}

\caption{\label{Fig1} EMT densities in the BB-model \cite{Varma:2020crx}.
(a) $T_{00}(r)$ (total) vs.\ $r$.
(b) $p(r)$ and $s(r)$ (total) vs.\ $r$ in the region of smaller $r$
($r\lesssim 2\,{\rm fm}$).
(c) $r^4 p(r)$ and $r^4 s(r)$  at very large $r$ ($r\gtrsim 2\,{\rm fm}$), 
where we see the new features (the power $r^4$ is included to enhance the features).}
\end{center}
\end{figure*}

In this work, we will focus on 3D EMT densities which 
are well-defined concepts in the large-$N_c$ limit,
for nuclei \cite{Polyakov:2018zvc,Hudson:2017xug},
and of course in classical models
\cite{Varma:2020crx}.
For discussions of 2D densities we refer to 
\cite{Lorce:2018egm,Freese:2021czn,Panteleeva:2021iip}. 
In Fig.~\ref{Fig1}a we show the energy density
which yields the mass of the system when integrated 
over the volume. $T_{00}(r)$ is always positive.
The characteristic discontinuity at $r=R$
is due to dust particles which by the construction of 
the BB-model are confined within the radius $r\le R$.
The solutions to the field equations are static with
$V^\mu=(V_0,0,0,0)$ and analogous for the Coulomb field.
At $r>R$, only the fields contribute to the energy
density which decay exponentially like 
$\phi(r)\sim \frac1r\,e^{-m_Sr}$ and 
$V_0(r)\sim \frac1r\,e^{-m_Vr}$ for $r\gg R$,
while the Coulomb potential is $A_0(r)\sim\frac1r$
for $r>R$.
The exact expressions for the fields and dust
particle distribution can be found in
\cite{Varma:2020crx,Bialynicki-Birula:1993shm}.

The pressure $p(r)$ and shear force $s(r)$ are defined 
through the components of the stress tensor, i.e.\ the 
$T_{ij}$ components of the EMT, as 
 \begin{equation}
  T^{ij}= \biggl( e_{r}^{i}e_{r}^{j}-\frac{1}{3}\delta ^{ij}\biggr)s(r)
       + p(r)\,  \delta ^{ij},\,
 \end{equation}
where $e_r^i$ is the unit vector in the radial direction. 
The total pressure, 
$p(r)=p_{\rm scal}(r) + p_{\rm vec}(r) + p_{\rm Coul}(r)$,
receives contributions from fields 
which are given by 
\begin{eqnarray} 
  p_{\rm scal}(r) &=& 
  -\frac16\,\phi'(r)^2-\frac12\,m_S^2\,\phi(r)^2,
  \label{Eq:p-scal}\\
  p_{\rm vec}(r) &=&
  \phantom{-}\frac16\,V_0'(r)^2+\frac12\,m_V^2V_0(r)^2,
  \label{Eq:p-vec}\\
  p_{\rm Coul}(r) &=& 
  \phantom{-}\frac{1}{6}\,A_{0}'(r)^{2}. 
  \label{Eq:p-Coul}
\end{eqnarray}

As can be seen in Eqs.~(\ref{Eq:p-scal}--\ref{Eq:p-Coul}),
the  scalar meson contribution is always negative, which 
corresponds to attractive forces directed towards the inside. 
On the other hand, the contributions of the vector mesons and the Coulomb field are always positive, which corresponds to repulsive
forces directed towards the outside.
When we integrate 
$\int_0^\infty dr\,r^2p_i(r)$ we get $-10.916\,{\rm MeV}$ from the 
scalar fields, $10.891\,{\rm MeV}$ from the vector field,
and a miniscule $0.025\,{\rm MeV}$ from the Coulomb field.
This reflects that the proton is a bound state of strong
forces and the electromagnetic contribution plays a minor 
role. But no contribution, no matter how small, can be
neglected as these numbers must add up exactly to zero
and fulfill von Laue condition,
\begin{equation}
     \int_0^\infty dr\,r^2p(r) = 0,
\end{equation}
which shows that the internal forces balance each other
and is a necessary condition for mechanical stability 
\cite{Goeke:2007fp}. The von Laue condition is exactly
satisfied in the BB-model 
\cite{Varma:2020crx,Bialynicki-Birula:1993shm}.

The shear force is  
$s(r)=\phi '(r)^{2}-V_0'(r)^{2}-A_{0}'(r)^{2}$.
Notice that the dust particles do not contribute to $s(r)$ 
and $p(r)$. 
The pressure and shear force are not independent but connected
by $\frac23\,s'(r)+\frac2r\,s(r)+p(r)=0$ due to EMT conservation.
The model results are shown in Fig.~\ref{Fig1}b.
The pressure inside the proton 
obtained from this model is an order of magnitude smaller than 
in the chiral quark soliton model \cite{Goeke:2007fp} or that
inferred from experiment \cite{Burkert:2018bqq}. This is because 
the BB-model is based on ``residual nuclear forces'' which are about 
an order of magnitude weaker than the strong forces among quarks 
inside the proton. 

The results for $s(r)$ and $p(r)$ in Fig.~\ref{Fig1}b are
qualitatively very similar to what was found in other
theoretical studies
\cite{Goeke:2007fp,Goeke:2007fq,Cebulla:2007ei,Jung:2013bya,Kim:2012ts,Jung:2014jja,Mai:2012yc,Mai:2012cx,Cantara:2015sna,Gulamov:2015fya,Nugaev:2019vru,Hudson:2017xug,Hudson:2017oul,Shanahan:2018nnv,Shanahan:2018pib,Anikin:2019kwi,Neubelt:2019sou,Azizi:2019ytx,Panteleeva:2020ejw,Kim:2020nug,Chakrabarti:2020kdc,Owa:2021hnj}.
In order to see the impact of long-range forces,
it is necessary to look more closely at the region of large $r$
which we shall do in the next section.

\section{Effects of long-range forces on the EMT}

In previous studies of strongly interacting systems governed 
by short-range forces, three common features were observed. 
The first feature is that the shear force is always positive. 
The second feature is that the pressure has one node at some
point $r_{0}$ with $p(r) > 0$ when $r < r_{0}$ and the pressure 
is less than zero for $r > r_{0}$. 
This property arises from the fact that the pressure must have 
at least one node to satisfy the von Laue condition, and the 
ground state exhibits a single node. 
Finally, the combination of $\frac23\,s(r)+p(r)$, which is 
normal force per unit area, is always positive. 

The BB-model is different from other studies, as it includes 
long-range Coulomb forces. From the model expressions for
$T_{00}(r)$, $s(r)$ and $p(r)$, we obtain the 
long-distance behavior which holds numerically for 
$r\gtrsim 2\,{\rm fm}$,  
\begin{eqnarray}
	T_{00}(r)   = \frac12\;
	\frac{\alpha}{4\pi}\;\frac{\hbar c}{r^4} + \dots,\,
	\label{Eq:T00-asymp}\\
    s(r)        = -\,
    \frac{\alpha}{4\pi}\;\frac{\hbar c}{r^4} + \dots,\, 
    \label{Eq:s-asymp}\\
    p(r)        = \frac{1}{6}\;
    \frac{\alpha}{4\pi}\;\frac{\hbar c}{r^4} + \dots,\,
    \label{Eq:p-asymp}
\end{eqnarray}
where the dots indicate contributions from the strong fields 
which are exponentially suppressed, and $\alpha$ is the 
fine-structure constant. We observe that $T_{00}(r)$ 
is always greater than zero which is in agreement with all 
prior studies. Because of the  $\frac{1}{r^4}$ decay of 
$T_{00}(r)$, the total energy converges but the mean square 
radius of the energy density diverges. 

In Fig.~\ref{Fig1}b we saw that $s(r)$ is positive,
which agrees with prior studies. But this is true only up 
to about $2.1\,{\rm fm}$ at which point $s(r)$ changes 
sign as shown in Fig.~\ref{Fig1}c.
Similarly, the picture of the pressure in the BB-model 
in Fig.~\ref{Fig1}b agrees with observations in other
studies with $p(r)$ turning from positive to negative 
around $0.8\,{\rm fm}$. However, looking more closely
in the region of larger $r$ we see that $p(r)$ exhibits
a second node around $2.4\,{\rm fm}$, and then remains 
positive. 
For completeness, we remark that the normal force,
$\frac23\,s(r)+p(r)$, exhibits an unusual feature and 
turns negative in the large $r$ region \cite{Varma:2020crx}.

In view of what has been learned from other studies
based on short-range forces, these three features are
counter-intuitive. It is an important observation that 
the presence of long-range interactions introduces new
features which have not been observed in prior studies
of EMT densities. One important practical implication
is the divergence of $D$-term which we shall review
in the next section.

\section{Divergence of the \boldmath $D$-term}

The $D$-term, ``the least known global property
\cite{Polyakov:2018zvc}'', is given
in terms of two equivalent definitions (arising from EMT
conservation) in terms of shear force and pressure,
\begin{equation}
    D   =  - \frac{4}{15}\,
          M \int d^3 r\; r^2 s(r) 
      =   M \int d^3 r\; r^2 p(r) \,. \label{eq:Dsp}
\end{equation}
The Coulomb contributions to $s(r)$ and $p(r)$ are minuscule
in the region $r<2\,{\rm fm}$, see Fig.~\ref{Fig1}b, 
giving the impression that the electromagnetic interaction
plays a very small role for the description of the structure
of a charged hadron. However small, the Coulomb contribution 
cannot be ignored, as it tells is that there is an electric 
charge. Especially at large $r$, the long-range 
$\frac1r$ behavior of the Coulomb contribution takes over
which has an important impact on the $D$-term.
Because of the asymptotic behavior of $s(r)$ and $p(r)$ 
at large $r$ in Eqs.~(\ref{Eq:s-asymp},~\ref{Eq:p-asymp}), 
both expressions for the $D$-term in (\ref{eq:Dsp}) diverge.
The fact that the $D$-term diverges due to long-range forces
is a new result, which has not been seen in prior studies. 

In order to obtain a finite (``regularized'') value for the 
$D$-term, one can introduce a regularization prescription. 
A unique regularization method can be derived by observing 
that, if the integrals were finite, then any linear combination 
of the two equivalent expressions in Eq.~(\ref{eq:Dsp}) would
give the same expression for $D$. However, the divergence can
be removed by considering one and only one linear combination 
which leads to finite regularized result for $D$, namely 
\begin{equation}
    D_{\rm reg} =  
      M \int d^3r\;r^2\biggl[\frac49\,s(r)+\frac83\,p(r)\biggr]
      \,.
\end{equation}
Numerically, we find $D_{\rm reg} = - 0.317$, i.e.\  
this regularization method preserves the negative sign 
of the $D$-term that has been observed in all prior studies.
The numerical value is about an order of magnitude smaller 
than \emph{e.g.}\ in the quark soliton model \cite{Goeke:2007fp}, 
which is expected as the BB-model is based on 
``residual nuclear forces'' that are weaker than the
strong interactions among quarks. It would be interesting 
to see if other methods exist to regularize these divergences. 

The form factor $D(t)$ in the BB-model is negative in a wide 
range of $t$. Only when
$(-t) \lesssim 2.8\times 10^{-4}\,{\rm GeV}^2$ does it become
positive, and diverges like $D(t)\sim 1/\sqrt{-t}$ for still
smaller $t$ \cite{Varma:2020crx}. Such small momentum transfers 
are currently beyond experimental reach. Noteworthy, the 
regularized value $D_{\rm reg}$ together with a quadrupole
fit, provide a very good approximation to the exact numerical
model results for $D(t)$ which confirms the practical
usefulness of the regularization method \cite{Varma:2020crx}.

\section{Model independent conclusions}

Our results for the EMT densities are model dependent in the 
region $r < \,$2-3$\,$fm, where the strong forces dominate. 
However, at $r \gg 3$ fm, exact QED calculations yield the 
same EMT density results as us, since QED has to reproduce
Maxwell's classical theory at long distances. 
In particular, the results in 
Eqs.~(\ref{Eq:T00-asymp},~\ref{Eq:s-asymp},~\ref{Eq:p-asymp}) 
are model independent and were obtained in QED
calculations \cite{Donoghue:2001qc,Metz:2021lqv}.
The divergence of $D(t)$ at small $t$ due to QED
effects was also found in chiral perturbation theory
calculations for charged pions \cite{Kubis:1999db}.
When comparing our results for $D(t)$ to those found 
using effective field theory techniques, we find that 
in the region $(-t) < 10^{-6}\,{\rm GeV}^2$, the model 
exactly reproduces QED
\cite{Donoghue:2001qc,Metz:2021lqv,Kubis:1999db}.

\section{Conclusion}
In Ref. \cite{Varma:2020crx}, we used a classical model 
\cite{Bialynicki-Birula:1993shm} which includes long-range
forces through the Coulomb contribution to calculate the
$D$-term. The classical character of the model was not an
impediment. It allowed us to investigate properties affected 
by the presence of long-range forces without worrying about 
the technical difficulties which arise when studying more
complicated quantum systems.
We found that the $D$-term of the proton diverges, in 
direct contrast to the convergent results of previous studies.
This feature is due to the infinite range of the electromagnetic interaction
and model independent. In fact, the model gives
$T_{00}(r)$, $s(r)$, $p(r)\sim \frac{\alpha}{r^4}$ at 
large $r$ \cite{Varma:2020crx} which agrees with 
QED calculations \cite{Donoghue:2001qc,Metz:2021lqv}.
In the model, we were able to derive a unique regularization
prescription to obtain a meaningful, finite, negative value
for the $D$-term in agreement with other studies. 
Without such a regularization, the form factor $D(t)$
changes sign and diverges at very small momentum
transfers below $-t < 10^{-4}\,{\rm GeV}^2$. 
While this $t$-region is currently out of reach
experimentally, it indicates that it may be necessary 
to refine the definitions of the EMT properties in the 
presence of long-range forces. It is currently an open
question how to do this in a model-independent way, or
whether the divergence of $D(t)$ may be remedied by  
considering QED radiative corrections.

\ \\
\noindent{\bf Acknowledgments.} 
 This work was supported by the NSF  
under the Contracts No.\ 1812423 and 2111490.


\section*{References}
\typeout{}
\bibliography{references}
\bibliographystyle{unsrt}

\end{multicols}
\end{document}